\documentclass[12pt]{article}

\pdfoutput=1

\usepackage[affil-it]{authblk}
\usepackage{graphicx}
\usepackage{amsmath,amssymb}
\usepackage{caption}
\usepackage[margin=1cm]{geometry}
\usepackage{color}
\usepackage[super,comma]{natbib}
\begin{document}

\title{Dynamics of thin liquid films on a porous substrate in zero gravity}

\author{Aneet Dharmavaram Narendranath%
  \thanks{Electronic address: \texttt{dnaneet@mtu.edu}; Corresponding author}}
\affil{Mechanical Engineering-Engineering Mechanics,\\ Michigan Technological University,\\ Houghton, MI 49931, USA}

%
\date{}

\maketitle

\section{Abstract}
The long-wave dynamics of liquid films on isothermal substrates show a dynamic competition between various physical mechanisms. If the destabilizing effect of thermocapillarity overcomes the stabilizing effect of surface tension and gravity, the liquid film ruptures in finite time, through the formation of primary and secondary thermocapillary finger structures.  

The long-wave evolution dynamics are compared for two different substrate types: isothermal non-porous and isothermal porous for small Biot number in a zero gravity environment. The multi-time-scale dynamics is revealed through time scales obtained from a method of similarity solutions.  It is observed that with an isothermal porous substrate, in zero gravity, secondary thermocapillary structures are damped through imbibition and that primary thermocapillary structures persist for long times without rupture.


\section{Background}

\begin{figure}
\centering
\includegraphics[width=0.6\textwidth]{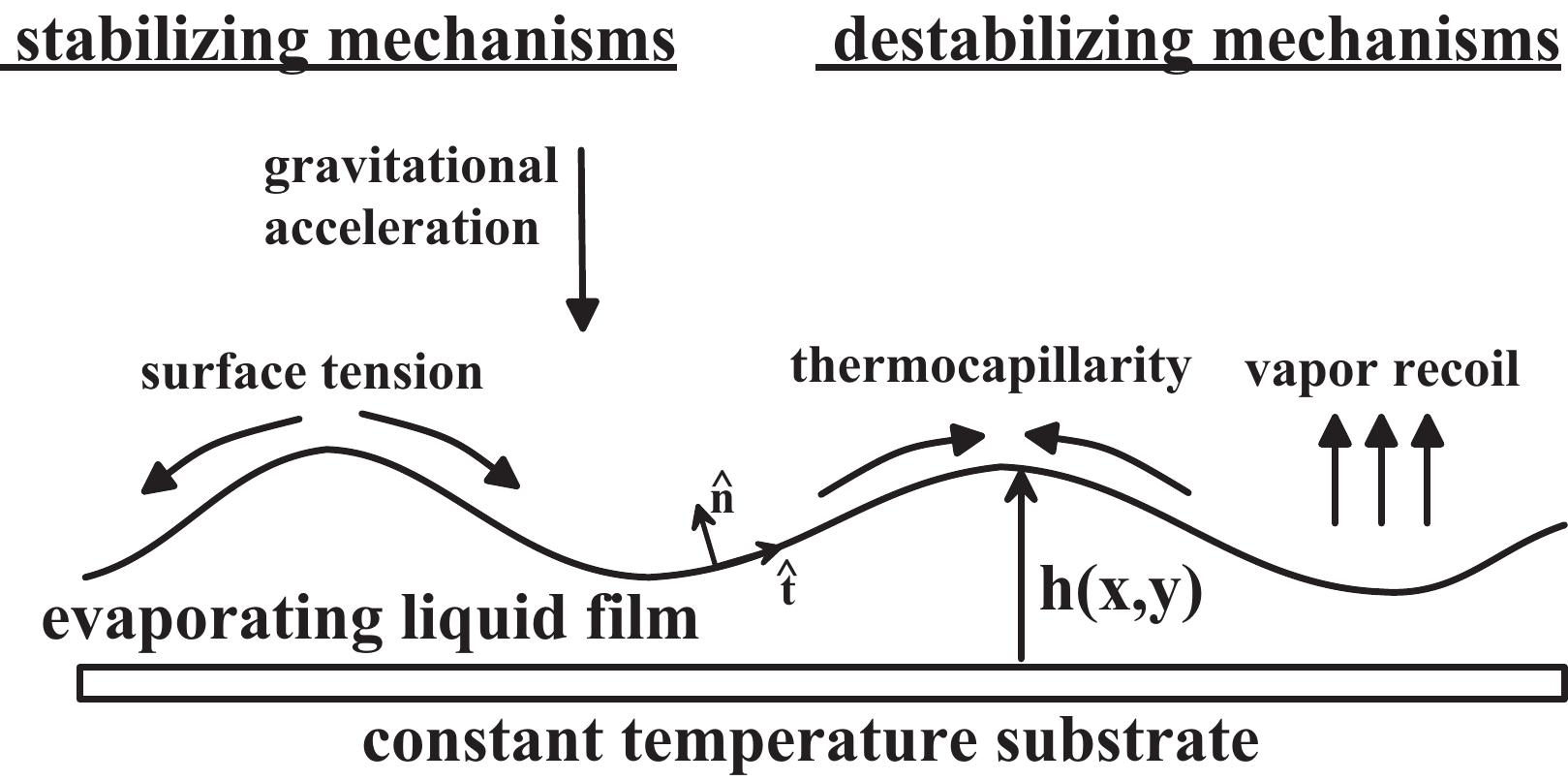}
\caption{Various stabilizing and destabilizing mechanisms that affect a liquid film on an isothermal solid substrate.  Periodic boundary conditions are used on the left and right edge.  In the present work, evaporation and concomitant vapor recoil effects are neglected.}
\label{fig00}
\end{figure}

Liquid film dynamics have been regularly studied using a non-linear evolution equation (equation \ref{e:evolution01}) \cite{Davis1982a,Burelbach,Oron2000a,Stone2002a}.  This is a stiff partial differential equation with non-linear terms (describing different stabilizing and destabilizing mechanisms) and non-dimensional numbers as multiplying coefficients or weights ($M,S,G,Bi,Pr$). Equation \ref{e:evolution01} describes a non-evaporating liquid film on an isothermal substrate under the influence of surface tension, gravity and thermocapillarity.  The direction of gravity is to stabilize any surface undulations.  A schematic of a liquid film on an isothermal solid substrate is depicted in figure \ref{fig00}.

The original evolution equation developed by \citet{Davis1982a} and \citet{Burelbach} included stabilizing effects of body forces (gravity), capillarity (surface tension) and destabilizing effects of thermocapillarity, evaporation, vapor recoil, and disjoining pressure.  The development of the evolution equation was based on viscous scaling, lubrication approximation, and normalization using the gravitational acceleration scale and surface tension scale.  Two-dimensional solutions of an evaporating liquid film were studied.  

\begin{table}
\centering
\caption{Nomenclature}
\label{t:nomenclature}

\begin{tabular}{cl}
$S = \sigma h_0/\rho \nu^2$ & Surface Tension number \\[5pt]
$G = g h_0^3/\nu^2$ & Galileo number \\[5pt]
$M = \sigma_T \Delta T h_0/\rho\nu\kappa$ & Marangoni number  \\[5pt]
$Pr = \nu/\kappa$ & Prandtl number \\[5pt]
$Bi$ & Biot number \\
$q$ & Wavenumber of instabilities\\
$q_\text{max}$ & Maximizing-wavenumber of instabilities\\
$\omega$ & Growth rate of instabilities \\
$\sigma$ & Surface tension \\
$h_0$ & mean film thickness \\
$\rho$ & mass density \\
$\nu$ & Kinematic viscosity \\
$\kappa$ & Thermal diffusivity \\
$k$ & Thermal conductivity \\
$\Lambda$ & Latent heat of evaporation\\
$X$ & Scaled length, $x_\text{physical}=X h_0 S$\\
$T$ & Scaled time, $t_\text{physical}=T \nu/(h_0^2 S)$\\
\end{tabular}
\end{table}

This equation allows for the study of the influence of individual mechanisms by engaging required non-linear terms through non-zero dimensionless coefficients. In equation \ref{e:evolution01}, for linear thermocapillarity, $\gamma=1$.  The rupture of liquid films on isothermal substrates is governed by a balance between destabilizing thermocapillarity and stabilizing surface tension and gravity.  In the absence of gravity, long-wave instabilities always occur when thermocapillarity overwhelms surface tension \cite{Narendranath2014a}. 

\begin{equation}\label{e:evolution01}
H_T + \underbrace{S \left( H^3 H_{XXX}\right)_X}_\text{surface tension} - \underbrace{G \left(H^3H_X\right)_X}_\text{Gravity/hydrostatic} + \underbrace{\frac{M Bi}{Pr}\left[\frac{H^2 \gamma}{(1 + Bi H)^2}H_X\right]_X}_\text{Thermocapillarity (T.C.)}= 0
\end{equation}

The non-linear evolution equation has a multi-time-scale nature.  The method of similarity solution \cite{Huppert1982a, Stone2002a, Chaudhury2015a, Witelski2015a} allows for the uncovering of the various time scales that affect liquid film dynamics. These time scales provide clues to the strength and rate of the various stabilizing and destabilizing and hence persistence of structures.  

Following the model put forth by Davis and Hocking \cite{Davis1999a, Davis2000a}, the effect of an isothermal porous substrate on zero-gravity evolution of non-evaporating liquid films is studied.  Effect of disjoining pressures is not included in the current work. In subsequent sections, the focus is on the effect of the substrate type on film dynamics. There exist other accounts in literature that study the effect of spreading of an axisymmetric droplet under the effect of injection or suction of fluid through a slot \cite{Momoniat2010a} or the dynamic control of falling films through injection and suction \cite{Thompson2016a}.  These did not take thermocapillary driven instabilities into account.  The current work studies the effect of zero gravity and a porous substrate on thermocapillary instabilities. 

%

\section{Evolution equation with isothermal porous substrate}
The modification of equation \ref{e:evolution01} to include an isothermal porous substrate yields equation \ref{e:evolution02}.  

\begin{equation}\label{e:evolution02}
H_T + \underbrace{S \left( H^3 H_{XXX}\right)_X}_\text{surface tension} - \underbrace{G \left(H^3H_X\right)_X}_\text{Gravity/hydrostatic} + \underbrace{\frac{M Bi}{Pr}\left[\frac{H^2 \gamma}{(1 + Bi H)^2}H_X\right]_X}_\text{Thermocapillarity (T.C.)}= \underbrace{\mathcal{S_W}N_{pm}\left(H_{XX} - G H\right)}_\text{Porous medium}
\end{equation} where $\mathcal{S_W}$ is a ``switch'' to turn the porosity term, $N_{pm}\left(H_{XX} - G H\right)$, on or off.  $N_{pm}$ is the non-dimensional Permeability Number, $\mathcal{K} \sigma/\mu$, where $\mathcal{K}$ is the ``permeability parameter'' from Darcy's law.

For linear thermocapillarity:
\begin{equation}\label{e:evolution03}
H_T + \underbrace{S \left( H^3 H_{XXX}\right)_X}_\text{surface tension} - \underbrace{G \left(H^3H_X\right)_X}_\text{Gravity/hydrostatic} + \underbrace{\frac{M Bi}{Pr}\left[\frac{H^2}{(1 + Bi H)^2}H_X\right]_X}_\text{Thermocapillarity (T.C.)}= \underbrace{\mathcal{S_W}N_{pm}\left(H_{XX} - G H\right)}_\text{Porous medium}
\end{equation}

In the analysis that follows, the Biot number, $Bi$ is chosen to be small ($Bi<1$) while the $M Bi/Pr \sim \mathcal{O}(1)$. For small Biot numbers, the temperature of the air-liquid interface is identical to the surface thermal profile.

\section{Results}
The film dynamics are studied after perturbing a quiescent liquid film with a sinusoidal perturbation with an amplitude several orders of magnitude less than the initial film thickness. To do this, the non-linear evolution equation is solved using the Wolfram \textit{Mathematica} symbolic solver.  This solver has proven accuracy in simulating the dynamics of evaporating and non-evaporating liquid films \cite{Narendranath2014a}. The method of lines along with the Livermore Solver for Ordinary Differential Equations (LSODA) for stiff equation is utilized for the solution.  The solution is terminated when the stiffness related errors cannot be resolved fruitfully. Termination time is called the film rupture time, $t_\text{rup}$.  At rupture, the lowest point in the film is less than several thousandths of the film's initial thickness.  The present numeric code has been validated against results in literature.

The liquid film fluid properties used were similar to low viscosity silicon oil, though no actual fluid could be found that matched these properties. The fluid corresponding to set of properties used by Williams and Davis\cite{Davis1982a} and \citet{Burelbach} and subsequent prominent researchers \cite{Oron2000a} will be referred to as the mathematical fluid.

\begin{table}
\centering
\begin{tabular}{l|l}
\textbf{Non-dimensional number} & \textbf{Value}\\
\hline
Gravity, G & 1/3 or 0\\
Surface tension, S & 100\\
Thermocapillarity, M & 5\\
Biot number, $Bi$ & 0.1\\
Permeability, $N_\text{pm}$ & zero or non-zero
\end{tabular}
\caption{Properties of the mathematical fluid}
\label{t:t1}
\end{table}

\subsection{Effect of substrate with non-zero gravity}

\begin{figure}
\centering
\includegraphics[width=0.7\textwidth]{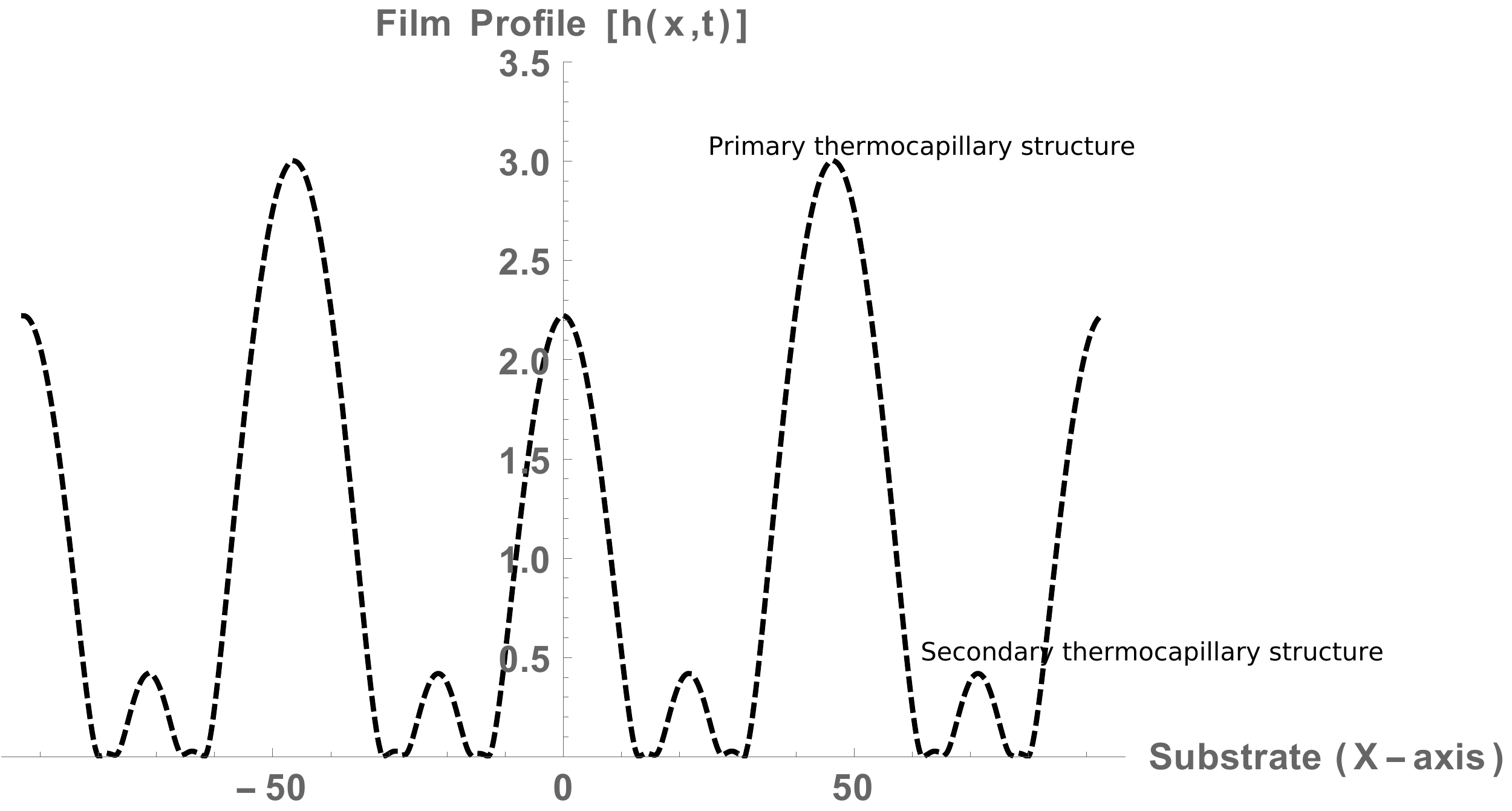}
\caption{Primary and secondary thermocapillary structures: Mathematical fluid with $G=1/3, S=100, M=5$ on an isothermal non-porous substrate $N_\text{pm}=0$}
\label{fig032}
\end{figure}

A non-zero gravity environment is first simulated to provide a comparison with zero-gravity film dynamics.  A thin film of the mathematical fluid placed on an isothermal non-porous substrate is depicted in figure \ref{fig032}.  Rupture takes place through the overwhelming of stabilizing effects of gravity and surface tension by the destabilizing effect of thermocapillarity.  Primary and secondary thermocapillary structures are formed during the evolution of the film profile.


\begin{figure}
\centering
\includegraphics[width=0.7\textwidth]{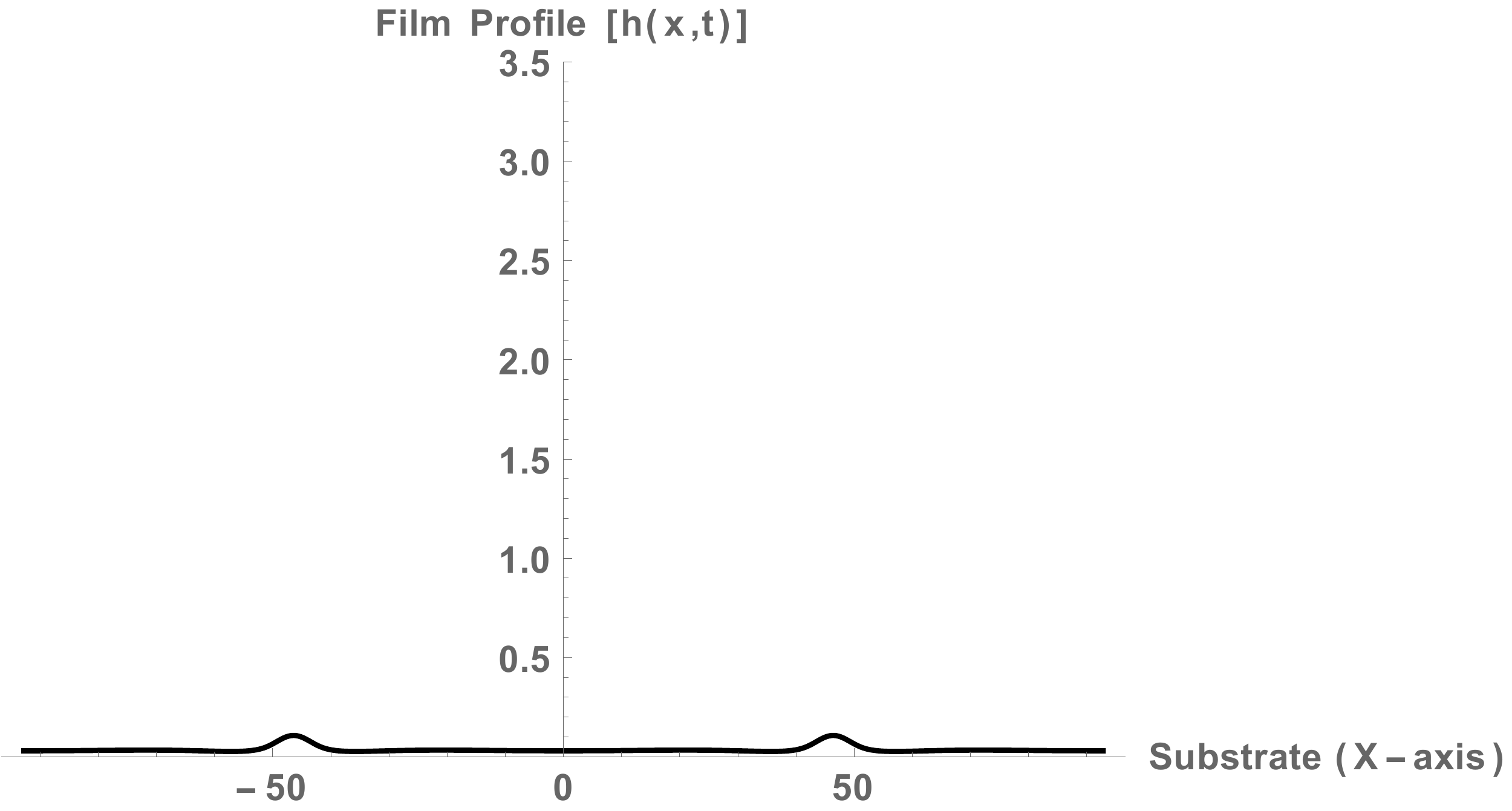}
\caption{Mathematical fluid with $G=1/3, S=100, M=5$ on an isothermal porous substrate $N_\text{pm}=0.025$}
\label{fig02}
\end{figure}

A thin film of the mathematical fluid placed on an isothermal porous substrate is depicted in figure \ref{fig02}.  In contrast with figure \ref{fig032}, the liquid film is depleted in finite time, through the porous substrate by the effect of capillary pressure ($H_{XX}$) and hydrostatic pressure ($G\,H$).

\subsection{Effect of substrate with zero gravity}

\begin{figure}
\centering
\includegraphics[width=0.7\textwidth]{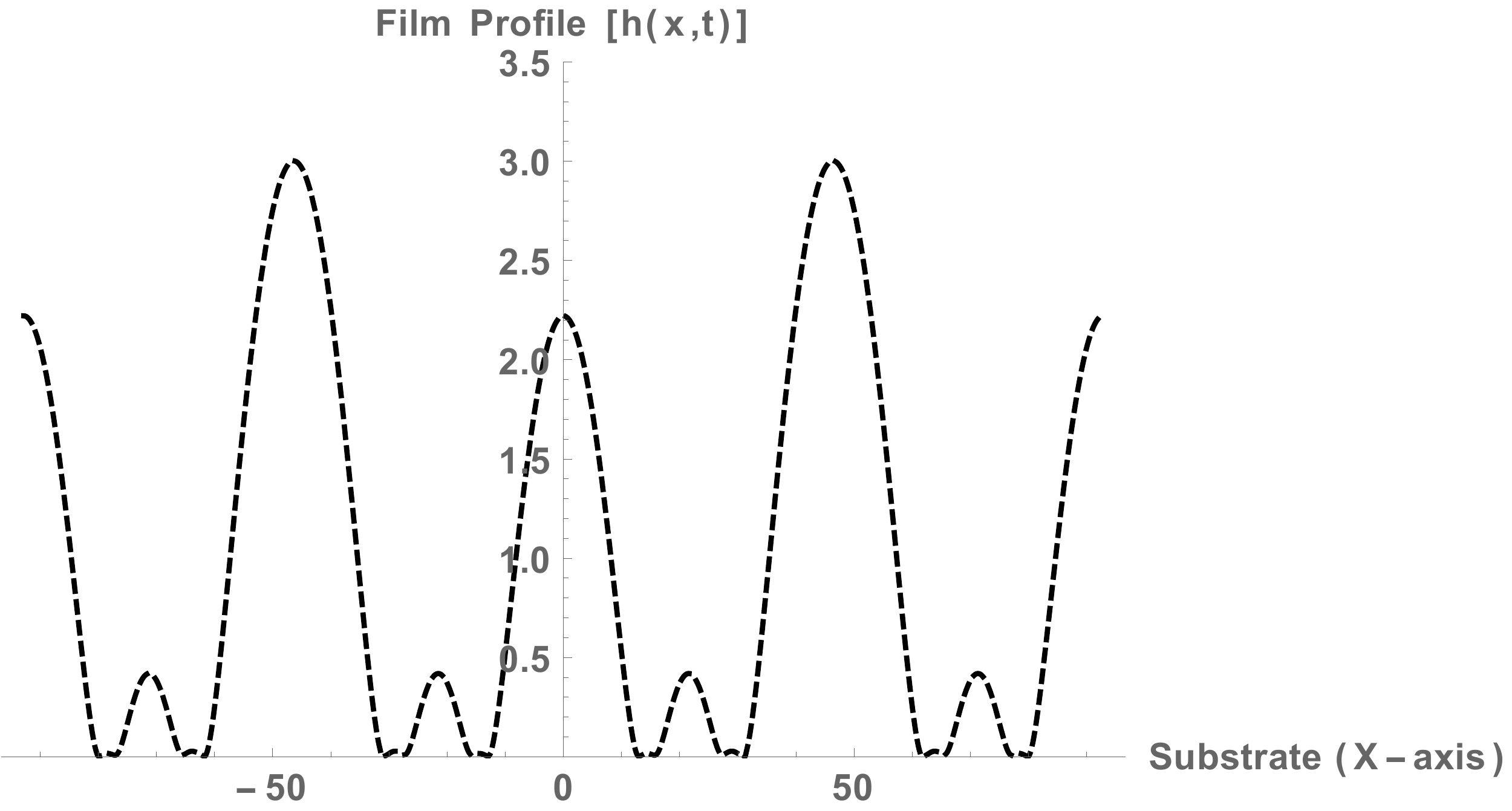}
\caption{Mathematical fluid with $G=0, S=100, M=5$ on an isothermal non-porous substrate $N_\text{pm}=0$}
\label{fig03}
\end{figure}

\begin{figure}
\centering
\includegraphics[width=0.7\textwidth]{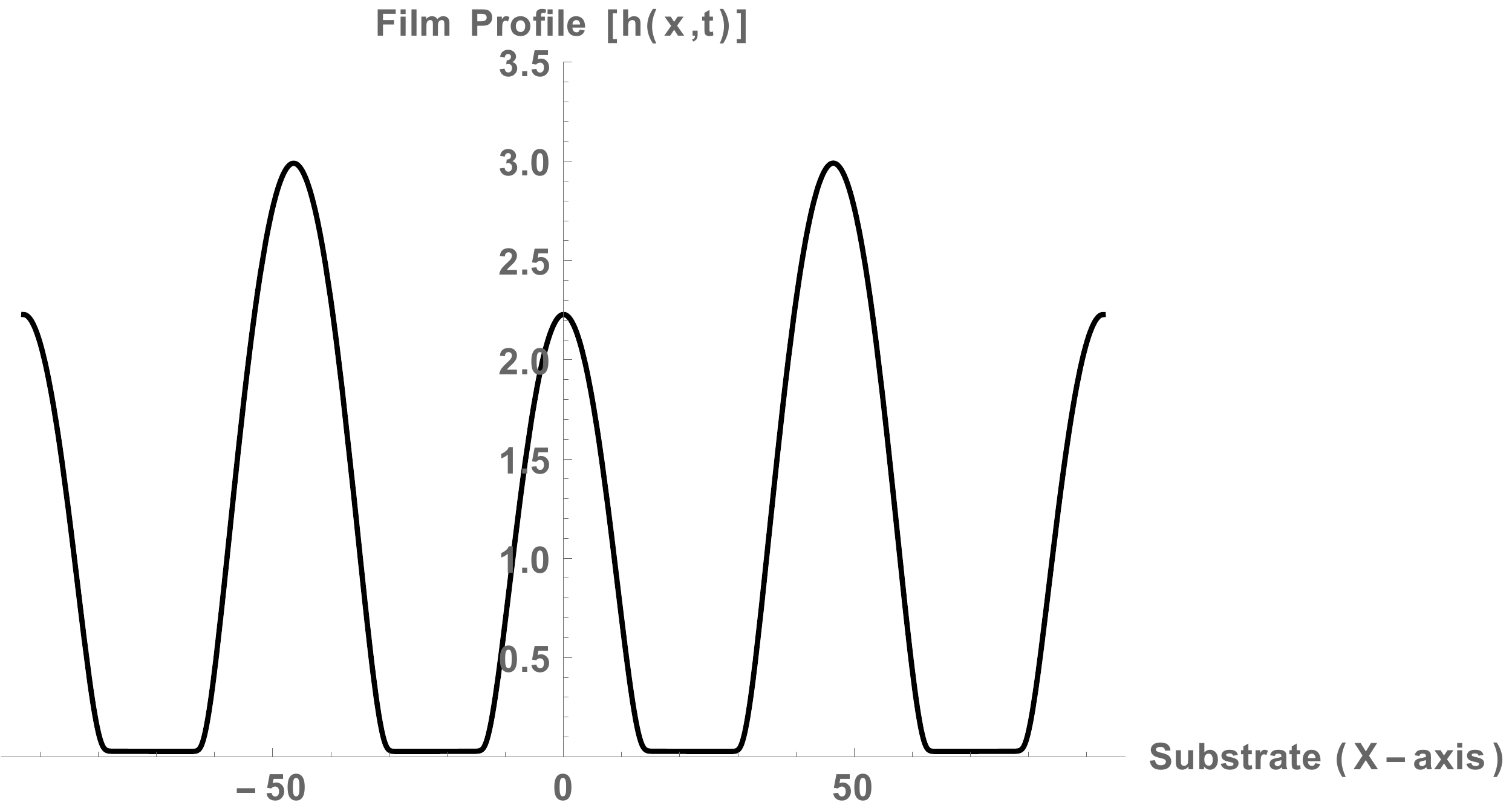}
\caption{Mathematical fluid with $G=0, S=100, M=5$ on an isothermal porous substrate $N_\text{pm}=0.025$}
\label{fig04}
\end{figure}

When the thin mathematical fluid film is placed in an environment with zero gravity, the isothermal non-porous substrate allows for finite time rupture at $t_\text{rup}$ as shown in figure \ref{fig03} while the isothermal porous film shows the presence of thermocapillary fingers that sustain for a long time ($\approx 32\,t_\text{rup}$) without rupture, as shown in figure \ref{fig04}.  Currently only one case of porosity with $N_\text{pm}=0.025$ is tested.

\subsection{Comparison of effect of substrate in zero gravity}
\begin{figure}
\centering
\includegraphics[width=0.7\textwidth]{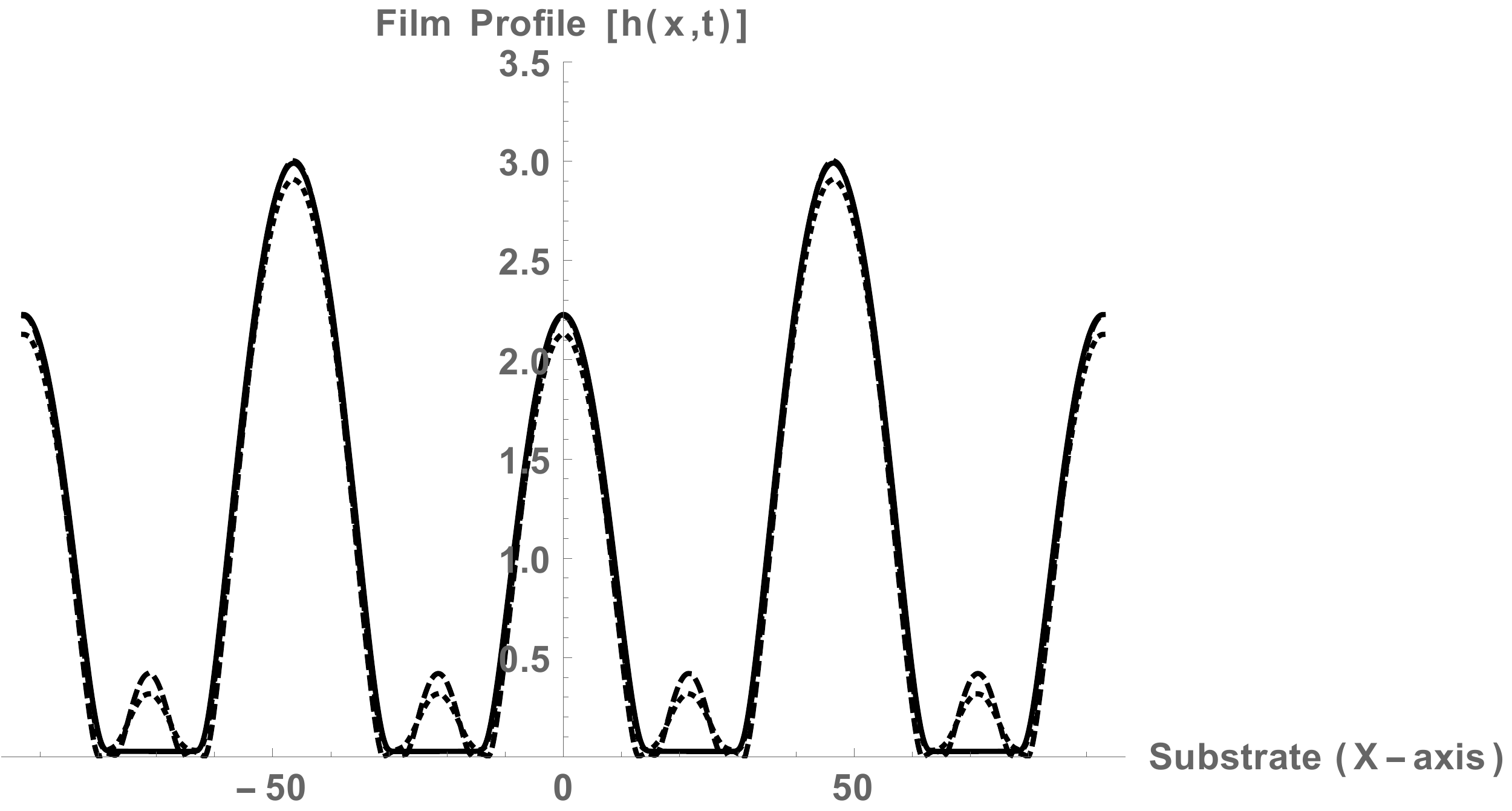}
\caption{Mathematical fluid with $G=0, S=100, M=5$ on an isothermal non-porous substrate ($N_\text{pm}=0$, dashed line).  Rupture takes place at $t_\text{rup}=551$ and an isothermal porous substrate ($N_\text{pm}=0.025$, dotted line at $t_\text{rup}$ and solid line at $32\,t_\text{rup}$).}
\label{fig05}
\end{figure}

The film profile evolution of the mathematical fluid film in zero gravity is compared for an isothermal porous substrate with that of an isothermal non-porous substrate, in figure \ref{fig05}. Thermocapillary finger structures (primary and secondary) form without and with an isothermal porous substrate.  In the case with a porous substrate, the secondary structures are damped and the primary structures persists for a long time ($32\times t_\text{rup}$) in the case with a porous substrate.  Here, $t_\text{rup}$ is the rupture time of the film with the non-porous substrate.  As a representative measure, the non-dimensional rupture time, $t_\text{rup}=551.0$ for the liquid film placed on the non-porous substrate.

\begin{figure}
\centering
\includegraphics[width=0.7\textwidth]{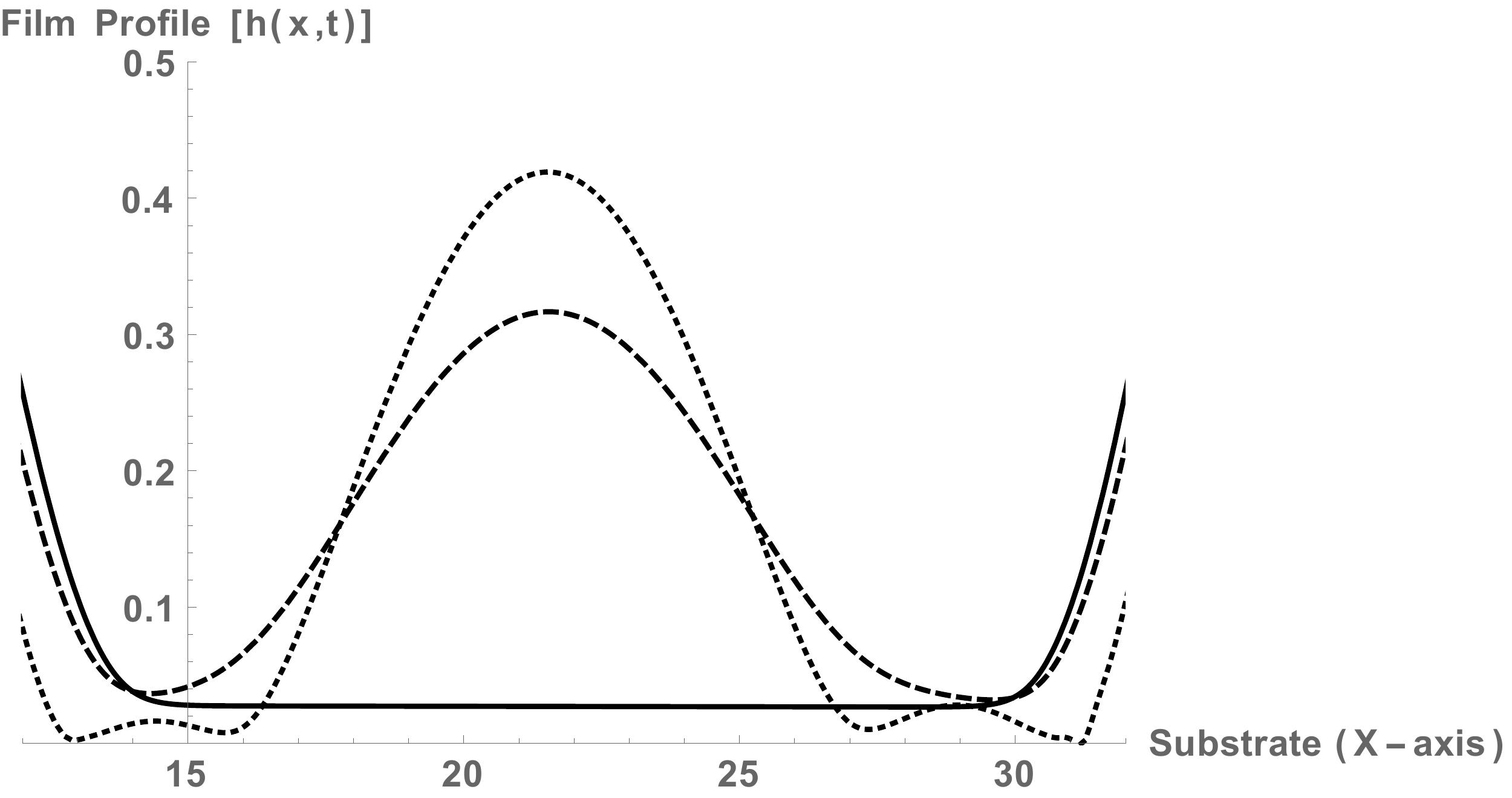}
\caption{Secondary structure dynamics of the mathematical fluid with $G=0, S=100, M=5$. The isothermal non-porous substrate ($N_\text{pm}=0$) driven dynamics are reported by dotted line.  Rupture takes place at $t_\text{rup}=551$.  The isothermal non-porous substrate ($N_\text{pm}=0.025$) driven dynamics are reported by a dashed line at $t=t_\text{rup}$ and a thick line at $t=32\,t_\text{rup}$.}
\label{fig06}
\end{figure}

A comparison of secondary structure evolution is depicted in figure \ref{fig06} with the non-porous (dotted line) and the porous substrates (dashed and thick lines, at different time instances). 

\subsection{Time scales}
Each of the mechanisms described by the evolution equation \ref{e:evolution01} has a certain time scale associated with it.  Using the method of similarity solution, the time scales are obtained and have been described in equation \ref{e:evolution04}.  In the obtaining of these time scales, the non-dimensional numbers, $S, G, M, Bi, Pr, N_\text{pm}$ have all been set to $\mathcal{O}(1)$.

\begin{equation}\label{e:evolution04}
\underbrace{H_T}_{H\sim T} + \underbrace{\left( H^3 H_{XXX}\right)_X}_{H\sim T^{-1/7}} - \underbrace{ \left(H^3H_X\right)_X}_{H\sim T^{-1/5}} + \underbrace{\left(H^2H_X\right)_X}_{H\sim T^{-1/4}}= \left( \underbrace{H_{XX}}_{H\sim T^{-1/2}} - \underbrace{H}_{H\sim T^{-1}}\right)
\end{equation}

\begin{figure}
\centering
\includegraphics[width=0.7\textwidth]{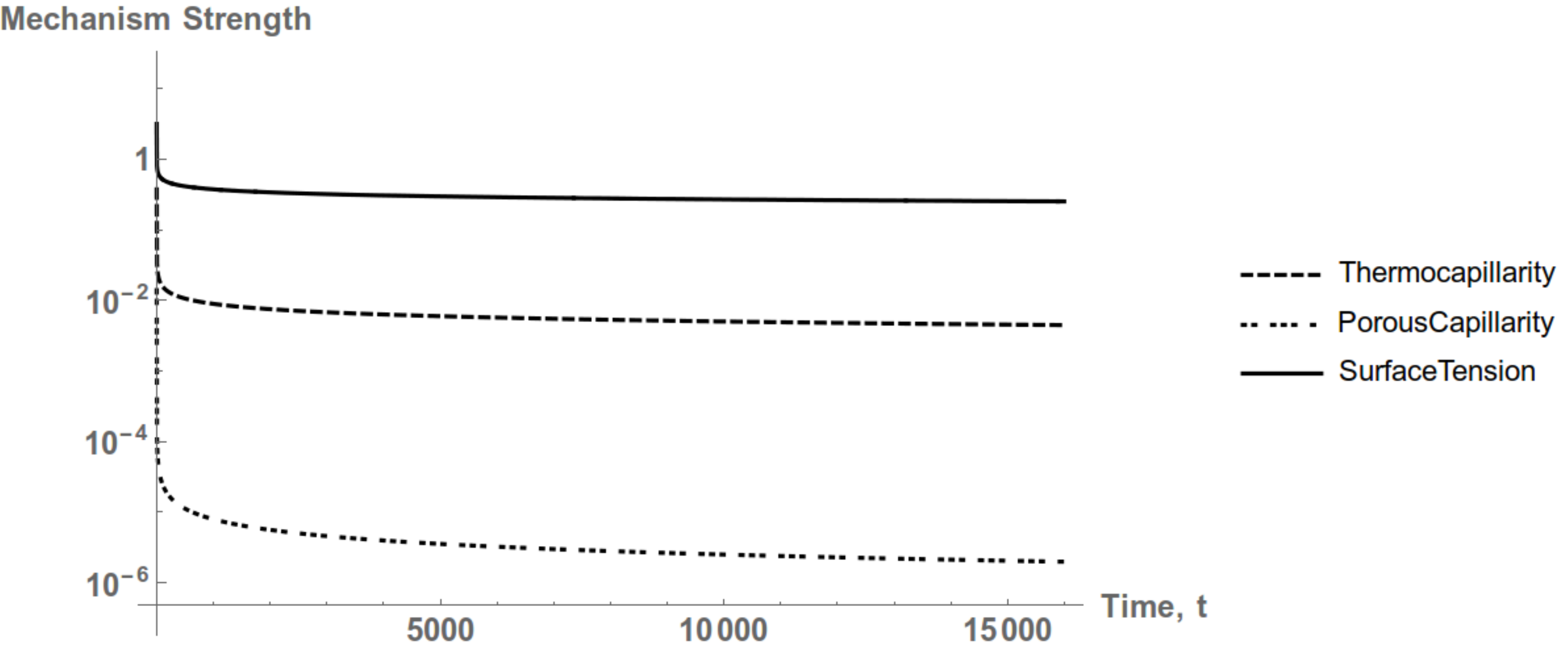}
\caption{Comparison of strengths of thermocapillarity, surface tension and capillarity due to porosity (PorousCapilarity)}
\label{fig07}
\end{figure}

At zero gravity $(G=0)$, comparison of the strengths of thermocapillarity, porosity (porosity driven capillarity) and surface tension reveal, in figure \ref{fig07}, that surface tension is the strongest term and has a stabilizing effect while thermocapillarity is the strongest destabilizing effect.  Capillarity induced by porosity is weaker than both surface tension and thermocapillarity but has a stabilizing effect nevertheless.

\begin{figure}
\centering
\includegraphics[width=0.7\textwidth]{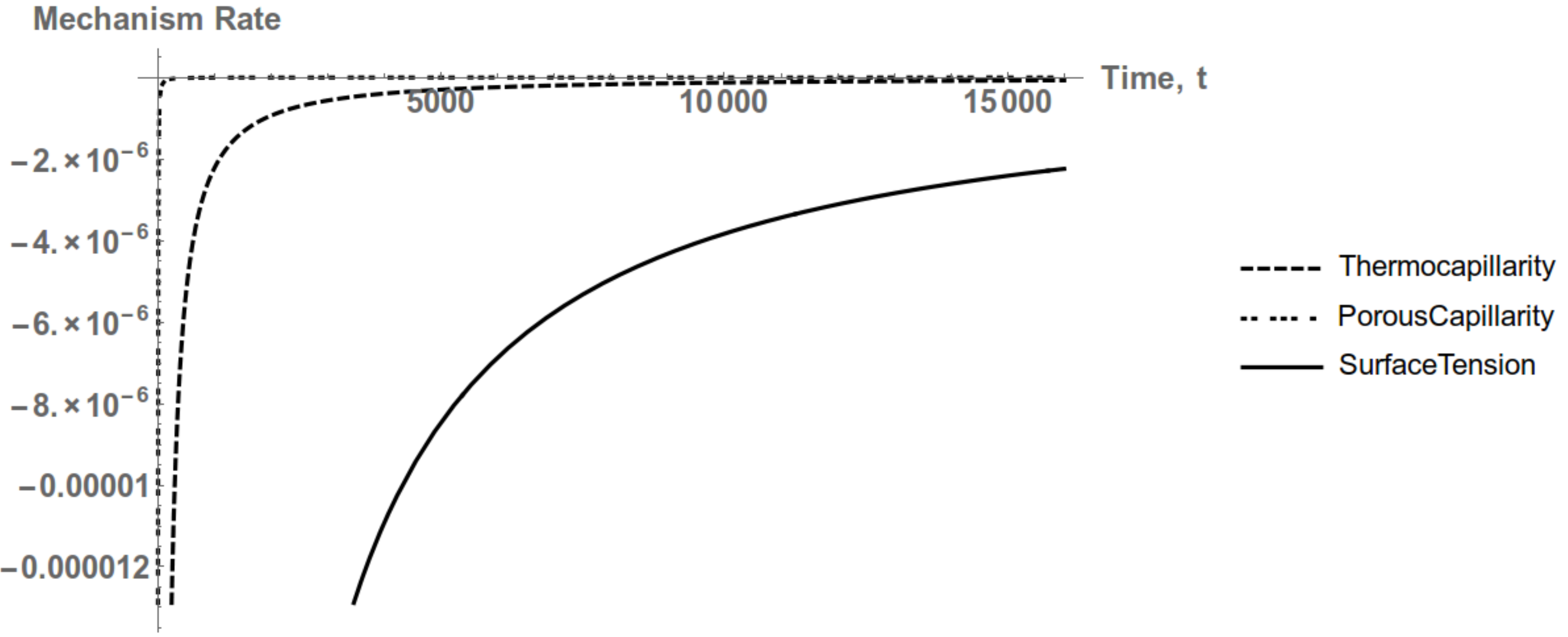}
\caption{Comparison of rates of thermocapillarity, surface tension and capillarity due to porosity (PorousCapilarity)}
\label{fig08}
\end{figure}

The rates of the thermocapillarity, surface tension and porosity are compared in figure \ref{fig08}, again at zero gravity.  As time progresses, stabilization due to porous capillarity matches that of destabilization rate of thermocapillarity. Surface tension has a weaker rate as compared to both thermocapillarity and porosity driven capillarity.  The rates of thermocapillarity and porosity driven capillarity tend to an asymptotic value at long time.  This signifies that at long time, a balance is struck between the various mechanisms and the liquid film has attained a likely static state.  Surface tension has the slowest rate and at time, $T=32\,t_\text{rup}$, it has not yet reached an asymptotic value. 

\subsection{Linear stability analysis}

A linear stability analysis provides an equation for the maximizing wavenumber ($q_\text{max}$) of long wave instabilities is shown in equation \ref{e:linstab}. In zero gravity ($Bo=0$), for $\frac{Bi M}{(Bi+1)^2 S}>\frac{Pm}{S}$, long wave thermocapillary (finger) structures form and persist.

\begin{equation}\label{e:linstab}
q_\text{max} = \frac{1}{\sqrt{2}}\left[\frac{Bi M}{(Bi+1)^2 S}-\text{Bo}-\frac{Pm}{S}\right]^{1/2}
\end{equation}


\section{Conclusions}
In this paper, the effect of an isothermal porous substrate on zero-gravity liquid film dynamics is studied.  A non-linear evolution equation for the film profile, modified to include a ``porous substrate term'' is utilized.  The fluid properties used are that of a mathematical fluid. Small Biot number ($Bi=0.1$) is used. The film dynamics are studied after perturbing a quiescent liquid film with a sinusoidal perturbation with an amplitude several orders of magnitude less than the initial film thickness.

Initially, it is observed that a porous substrate with a non-zero gravity environment can have a stabilizing effect on long wavelength instabilities (thermocapillary finger structures) through drainage of the film through the substrate. In zero gravity, a porous substrate can lead to the formation of thermocapillary finger structures that persist for long periods of time. This could potentially lead to using substrate porosity as a control parameter to stabilize unstable liquid films or to impose long lasting thermocapillary structures on the substrate. 

In a zero gravity environment, a porous substrate leads to the drainage and damping of secondary thermocapillary structures.  A comparison of dynamics at different instances in time reveals that there is barely any damping of primary structures (as compared to a case with non-porous substrate) but it is only the secondary structures that are drained through the capillary effect of the porous substrate.  A porous substrate in a zero gravity environment has the effect of linearizing non-linear effects.  In other words, the second order linear $H_{XX}$ term can overwhelm non-linear destabilisation effects due to thermocapillarity $\left(\frac{M Bi}{Pr}\left[\frac{H^2}{(1 + Bi H)^2}H_X\right]_X\right)$.

The method of similarity solutions is used to obtain the time scales of various stabilizing and destabilizing mechanisms.  In zero gravity, it is observed that the porosity driven capillarity has a stronger stabilizing rate than the destabilizing rate of thermocapillarity.  Surface tension has the slowest rate of the three.

A linear stability analysis is performed to derive an equation for the maximizing wavenumber. A condition put forth for the emergence of long wave instabilities for a liquid film in zero gravity on an isothermal porous substrate.

\bibliographystyle{unsrtnat}
\bibliography{ref}

\end{document}